**Raman spectroscopy of wurtzite and zinc-blende GaAs nanowires: polarization dependence, selection rules and strain effects**.

I. Zardo,[1] S. Conesa-Boj,[2] F. Peiro,[2] J.R. Morante,[2,3] J. Arbiol,[2,4] E. Uccelli,[1,5] G. Abstreiter,[1] A. Fontcuberta i Morral[1,5]

[1] Walter Schottky Institut and Physik Department, Technische Universität München, Am Coulombwall 3, 85748 Garching (Germany)

[2] Departament d'Electronica, Universitat de Barcelona, 08028 Barcelona, CAT, Spain

[3] IREC, Catalonia Institute for Energy Research, Barcelona 08019, Spain

[4] ICREA Research Professor at Institut de Ciència de Materials de Barcelona, CSIC, 08193 Bellaterra, CAT, Spain

[5] Laboratoire des Matériaux Semiconducteurs, Institut des Matériaux, Ecole Polytechnique Fédérale de Lausanne, CH-1015 Lausanne, Switzerland

Polarization dependent Raman scattering experiments realized on single GaAs nanowires with different percentages of zinc-blende and wurtzite structure are presented. The selection rules for the special case of nanowires are found and discussed. In the case of zinc-blende, the transversal optical mode $E_1(TO)$ at 267 cm$^{-1}$ exhibits the highest intensity when the incident and analyzed polarization are parallel to the nanowire axis. This is a consequence of the nanowire geometry and dielectric mismatch with the environment, and in quite good agreement with the Raman selection rules. We also find a consistent splitting of 1 cm$^{-1}$ of the $E_1(TO)$. The transversal optical mode related to the wurtzite structure, $E_2^H$, is measured between 254 and 256 cm$^{-1}$, depending on the wurtzite content. The azimuthal dependence of $E_2^H$ indicates that the mode is excited with the highest efficiency when the incident and analyzed polarization are perpendicular to the nanowire axis, in agreement with the selection rules. The presence of strain between wurtzite and zinc-blende is analyzed by the relative shift of the $E_1(TO)$ and $E_2^H$ modes. Finally, the influence of the surface roughness in the intensity of the longitudinal optical mode on {110} facets is presented.

1. **Introduction**

Nanowires are filamentary crystals with a large aspect ratio and diameters in the range of several to several tens of nanometers. Due to their small diameters, large surface-to-volume ratios, and novel electronic and optical properties, nanowires are objects of intense study in fundamental science [1,2,3,4,5]. One of the consequences related to the large surface-to-volume ratio is the crystallization of nanowires in crystalline structures that are not stable in the bulk form [6,7,8,9,10,11]. In the case of nanowires pertaining to the arsenides and phosphides, it is very common to find rotational twins along with polytypism between wurtzite and zinc blende structures. Wurtzite is not stable in the bulk form of these compounds. The occurrence of this phase as one of the main crystalline structure of arsenide and phosphide nanowires has been discussed by several authors. From the thermodynamic point of view, the surface energy of {1100} wurtzite planes should be lower than the {110} and {111}A/B of zinc-blende [12]. This means that for small radii nanowires, the wurtzite phase should be more stable than the zinc-blende. In the case of GaAs, the critical radius under which wurtzite is expected to be the most stable phase has been predicted to lie between 5 and 25.5 nm [13,14,15].

Raman scattering, or more generally inelastic light scattering is a standard non-destructive, contactless characterization technique of materials which allows to access mainly the phonon modes at the Γ point and in some cases to the dispersion [16,17,18]. Raman spectroscopy can be realized by using a confocal microscope, thereby obtaining lateral submicron resolutions of the properties of a material. It has been developed to a versatile tool for the characterization of semiconductors leading to detailed information on crystal structure, phonon dispersion, electronic states, composition, strain and so-on of semiconductor nanostructures or nanowires [19,20,21,22,23,24,25,26,27]. More recently Raman spectroscopy has been used extensively to study also nanowires and quantum dots [28,29]. Several new phenomena have been reported to date with respect to one-dimensional structures. Polarization dependent confocal Raman spectroscopy on single carbon nanotubes and/or nanowires indicates that the physics behind Raman scattering of such one-dimensional nanostructures can differ significantly from the bulk. Indeed, the highly anisotropic shape of the nanowires can lead to angular dependencies of the modes which otherwise would not be expected from the selection rules [30,31,32]. Additionally, new modes such as the ones associated with surface phonons have been detected [33,34,35,36].

In this work, scanning Raman spectroscopy is performed on GaAs nanowires with zinc-blende and wurtzite structures. To the best of our knowledge, no Raman results have been reported so far on GaAs nanowires with wurtzite structure or with a mixture of wurtzite and zinc-blende. The paper is structured as follows: After presenting theoretical considerations on the phonon dispersion and selection rules for both structures in Section 2, the experimental details concerning the polarization dependent Raman spectroscopy are presented in Section 3. In Section 4 the selection rules, effect of strain and surface roughness on the Raman spectra are discussed. Finally, a conclusion is given in Section 5.

## 2. Theoretical considerations on the Raman scattering of zinc-blende and wurtzite GaAs.

In Raman scattering experiments the scattering intensities can be calculated from the Raman tensor which depends on the crystal symmetry:

$$I_s \prec \sum \left| \vec{e}_i \cdot R \cdot \vec{e}_s \right|^2 \qquad (1)$$

where **R** is the Raman tensor and $\vec{e}_i$ and $\vec{e}_s$ are the polarization of the incoming and scattered light respectively. In zinc-blende GaAs the TO and LO modes are split at k = 0 due to the polar nature of the crystal. The so-called E1 (TO) mode is allowed in backscattering from the (110) and (111) surfaces while the A1 (LO) mode is allowed in backscattering from (100) and (111) surfaces, respectively.

For the calculation of the selection rules the standard base is not appropriate, as one has to consider the crystal structure and morphology of the nanowire. The logical set of axes correspond to the directions x=[0-11], y=[211] and z=[-111]. A schematic drawing of a reference GaAs sample prepared for the backscattering measurements under this axis system is shown in Fig. 1a. The reference axes x, y and z are schematized in the figure. The x axis corresponds to the direction of incident and scattered light [0-11], while y and z are the in plane axes perpendicular and parallel to the nanowire axis, respectively [211] and [-111]. The values of the Raman tensor for the transversal modes in that configuration for incident light along the x axis are:

$$R'(y) = \begin{pmatrix} 0 & \frac{1}{\sqrt{3}} & -\frac{1}{\sqrt{6}} \\ \frac{1}{\sqrt{3}} & \frac{2}{3} & -\frac{1}{3\sqrt{2}} \\ -\frac{1}{\sqrt{6}} & \frac{1}{3\sqrt{2}} & -\frac{2}{3} \end{pmatrix}, \; R'(z) = \begin{pmatrix} 0 & -\frac{1}{\sqrt{3}} & \frac{1}{\sqrt{6}} \\ -\frac{1}{\sqrt{3}} & \frac{2}{3} & \frac{1}{3\sqrt{2}} \\ \frac{1}{\sqrt{6}} & \frac{1}{3\sqrt{2}} & -\frac{2}{3} \end{pmatrix} \qquad (2)$$

After transformation of the Raman tensor into the new basis, the selection rules are obtained by calculating the scattering cross-section using eq. (1) [37]. The predicted intensity of the scattered light polarized parallel or perpendicular to the [-111] direction, $I_s(\parallel)$ and $I_s(\perp)$, as a function of the angle $\alpha$ of the incident polarization with respect to the [-111] axis is plotted in Fig.1c. The maximum in $I_s(\parallel)$ is obtained for an angle of the incident polarization of -20°, while for $I_s(\perp)$ it is obtained for 70° with respect to the [-111] direction. Polarization dependent Raman scattering measurements were realized on a reference GaAs with the configuration described above. A parallel polarized Raman spectrum from a bulk GaAs (0-11) in backscattering geometry, obtained using exciting light polarized along the [-111] direction is shown in Fig. 1b. The spectrum is composed by the $E_1$(TO) phonon mode at 267.2 cm$^{-1}$, as expected. The peak exhibits a full width at half maximum (FWHM) of 2.8 cm$^{-1}$, indicating the high crystal quality. The azimuthal dependence of the scattered light is plotted in Fig. 1d. The experimental curve is very similar to the theoretical prediction. We notice a shift in the intensity maxima of about 10°, which we attribute to experimental error due to the limited accuracy in aligning the polarization of the exciting light with the crystallographic [-111] direction.

Wurtzite GaAs is not stable in the bulk form. It is found under special conditions such as high pressures and in the nanoscale form such as nanoparticles and nanowires [38,39,40,41,42,43]. To our knowledge, the phonon spectrum of wurtzite GaAs has not been calculated or measured yet. As it has been shown in the case of GaN and SiC, the main characteristics of the phonon dispersion of wurtzite GaAs can be deduced in an approximated way thanks to the straight forward relation between the two structures [44,45]. From the crystallographic point of view zinc-blende and wurtzite structures differ only in the stacking periodicity of the Ga-As bilayers parallel to the $(111)_{ZB}/(0001)_{WZ}$ oriented planes. For illustration, a schematic representation of the crystalline structure of zinc-blende and wurtzite structures is given in Fig. 2a and b. The notation 'a', 'b', 'c' denotes the position of the Ga-As pairs in the $(111)_{ZB}/(0001)_{WZ}$ planes. The stacking order is 'abcabc' for zinc-blende and 'abab' for wurtzite [46]. As the unit cell length of wurtzite along the [0001] axis is double with respect to the zinc-blende along the [111], the phonon dispersion can be approximated by folding the one for the zinc-blende structure along the [111] direction ($\Gamma \rightarrow L$). This operation is represented schematically in Fig. 2c. As a consequence of the folding for the case of wurtzite, four new modes appear at the $\Gamma$ point of the Brillouin zone. The folded modes are indicated with dashed lines in Fig. 2. Group theory also predicts nine phonon normal modes at the $\Gamma$ point and which modes are infrared and/or Raman active. In Table 1 the results of these considerations are presented. The notation used in the table is the standard Raman notation and it can be understood as follows: outside the bracket, the symbols show respectively from left to right the direction of incident and scattered light. Inside the bracket, they give from left to right the polarization direction of the incident and scattered light, respectively. We use the equivalent axes which were used in the case of zinc-blende [47]. The directions of the x, y and z axis correspond respectively to the crystallographic directions [11-20], [1-100] and [0001]. In Table 1, the expected energies of the wurtzite modes at the $\Gamma$ point are given. From the six modes, the two $B_1$ are silent modes non observable by neither Raman nor Infrared spectroscopy. The other four modes are Raman active and two of them infrared active at the same time $-A_1$ (LO) and $E_1$ (TO). In conclusion, four modes should be observable by Raman spectroscopy of wurtzite GaAs: (1) two already existing in zinc-blende: $A_1$ (LO) and $E_1$ (TO) and (2) two specific from the wurtzite phase: $E_2^H$ and $E_2^L$ at 259 and 59 $cm^{-1}$ respectively. Finally, there is an additional effect that one should consider to describe precisely the phonon dispersion of wurtzite GaAs. Indeed, due to the anisotropy of the atomic bonds along or in the perpendicular direction of the c axis, the $E_1$ and $A_1$ modes may mix, giving rise to two apparent new modes $E_1$(LO) and $A_1$(TO) [48,49]. To our knowledge, the exact position for GaAs has not been predicted or measured.

### 3. Experimental

Gallium arsenide nanowires were grown by molecular beam epitaxy by the gallium assisted method, which avoids the use of gold as a catalyst, as described elsewhere [50,51]. The growth precursors are atomic gallium and $As_4$ molecules. Gallium forms droplets at the surface and drives the nucleation of the nanowires. The growth rate of the nanowire is given by the $As_4$ flux [51]. The samples presented in this paper have been obtained at arsenic pressures between $3\times10^{-6}$ and $3\times10^{-7}$ mbar which represent growth rates respectively between 3 and 0.3 $\mu$m/h.  We have shown in the past that a reduction in the growth rate results in an increase in the fraction of wurtzite in the nanowire [52].

The structure of the nanowires was studied by high resolution transmission electron microscopy (HRTEM) in a Philips/FEI CM 300. Samples were prepared as follows for the TEM measurements: First, they were mechanically removed from the substrate with a razor blade and diluted in a hexane suspension. A drop was then deposited on a holey carbon copper grid. Before introducing the sample into the microscope, it was cleaned in a plasma cleaner for 15 s in order to remove any organic surface contaminants.

Raman spectroscopy was realized in the backscattering configuration on single GaAs nanowires transferred on to a silicon substrate.  Prior to the measurements, the nanowires were located by imaging the surface with a CCD. The 514.5 nm line of an $Ar^+$ laser was used for excitation. The power of the incident light was about 200 $\mu$W (equivalent to 70 $kW/cm^2$). Special care was taken to ensure that the nanowires were not heated [31,53]. A series of measurements with increasing the excitation power was realized in order to choose the excitation regime which was well below to the threshold under which heating effects were observed. The scattered light was collected by an XY Raman Dilor triple spectrometer with a multichannel charge couple device detector. The sample was positioned on a XY piezo-stage, which allowed the scanning of the surface (and therefore the nanowire) with a precision of 10 nm. The polarization dependent measurements were realized as follows. First, the incoming light passes through a λ/2 plate, so that its polarization $e_i$ can be rotated by an angle φ. After passing through a beam splitter (50:50), it is focused on the nanowire with a 100x objective (0.95 NA). The polarization of the scattered light $e_s$ is analyzed by measuring the intensity of the two components (parallel or perpendicular to the wire). For this, a second polarizer is used. The efficiency of the spectrometer depends on the polarization of the incoming light and is higher when it is perpendicular to the slit. In order to avoid artefacts linked to this, a λ/2 plate at the entrance of the spectrometer was added in order to flip the polarization of the light into the most efficient direction. A schematic drawing of the set-up and the configuration used for the measurements on the nanowires is shown in Fig. 3.

In order to obtain the phonon peak position, width and intensity, all measured spectra were fitted in the 200 - 300 cm$^{-1}$ frequency range with the sum of lorentzians and a background contribution.

## 4. Results

Two types of nanowires were investigated. The first sample consists of GaAs nanowires with 100% zinc-blende structure. The second type of nanowires corresponds to GaAs with 30% wurtzite phase [51]. The results are presented separately in the next two sections.

### 4. 1. Zinc blende GaAs nanowires

GaAs nanowires with 100% zinc-blende structure were obtained by growing them at a growth rate of about 3μm/h, as published elsewhere [54]. Representative HRTEM measurements of these nanowires are presented in Fig.4, where a bright field TEM image is shown. Stripes with different contrast are observed along the nanowires. Careful analysis indicates that those correspond to the presence of twins perpendicular to the growth axis. A detail of such defect is presented in Fig.4b. Each side of the twin corresponds to (110) and (011) facets [55,56], the growth axis remains parallel to the (111)B direction – or equivalently (-111) , C spot in Fig. 4c.

The nanowires exhibit a hexagonal section, with facets of the {110} family and a growth axis along the [-111] direction. Raman spectroscopy measurements were realized on nanowires transferred on a silicon substrate, so that the flat facet of the family {110} is perpendicular to the incident light, as schematized in Fig. 3. This means that the incident and scattered light are parallel to the [0-11] axis and the polarization is in the plane defined by the [-111] and [211] axes.

We start by presenting the measurements on a reference sample. One of the objectives of this study is to determine the selection rules of zinc-blende and wurtzite GaAs in the nanowire geometry. In order to distinguish between the effect of the crystalline structure and the nanowire anisotropy, we start by performing the experiments with zinc-blende GaAs in the bulk and nanowire form. The reference sample consists of a (-111) oriented GaAs wafer cleaved along the (0-11) facet. The measurements on the reference sample are presented in Fig.1b and d. The spectrum is composed by one unique phonon mode –$E_1$(TO)- at 267.2 cm$^{-1}$ as expected from the selection rules discussed in Section 2.

The configuration used for the bulk sample -shown in Fig. 1a- is equivalent to the one used for the nanowires – shown in Fig. 3. This enables us to compare the polarization dependence between the two samples. The Raman intensity seen in a scan along the nanowire axis is presented in Fig. 5a. This measurement has been realized by exciting with the polarization of incident light along the nanowire axis. The Raman scattering spectrum is composed mainly by the $E_1$(TO) mode at 266.7 cm$^{-1}$, slightly lower than for the reference sample. The $A_1$(LO) is not observed, as it is forbidden in backscattering geometry for surfaces of the family {110}. The spectrum is homogeneous along the nanowire. A slight intensity decrease is observed. This is an artifact of the measurement. The scan direction is not exactly parallel to the nanowire axis, therefore the nanowire gradually moves out of the measuring spot during the scan and the intensity decreases. The Raman spectra of a nanowire and the bulk reference sample – GaAs (0-11)- are compared in Fig. 5b. The measurements have been taken under the configuration $x(z,z)\bar{x}$, meaning that the exciting and scattered light are polarized along the [-111] direction, the nanowire growth axis. Both spectra have been normalized with respect to the intensity of the TO mode as absolute intensities can not be compared easily. The TO peak of the nanowires has a FWHM of 4.8 cm$^{-1}$, about 2 cm$^{-1}$ larger than the reference sample. This value indicates still an excellent crystal quality for the nanowires. The broadening of the peak with respect to the reference can be attributed to the presence of the twins, which introduce some disorder in the crystal structure.

The azimuthal dependence of the Raman scattering is shown in Fig.5c. There, the intensity of the scattered light parallel and perpendicular to the nanowire axis –I$_s$($\parallel$) and I$_s$($\perp$)- as a function of the angle of the incident light polarization with respect to the [-111] direction is plotted. In the measuring configuration $x(\alpha,z)\bar{x}$, the maximum intensity is obtained for $x(z,z)\bar{x}$: when incident and analyzed polarization are parallel. Some experimental points slightly deviate from the trend. We attribute it to an experimental artifact [57]. In the measuring configuration $x(\alpha,y)\bar{x}$, there is no preferential value of α giving a maximum scattering intensity. The scattered light polarized along the [211] direction, namely perpendicular to the wire, has a drop in the intensity by about a factor 6. We note that this effect is not observed for the reference sample, indicating that the selection rules in this case are altered. As it will be discussed in the following, we believe that this is most probably due to the one-dimensional geometry of the sample.

## 4. 2. Nanowires with 30% wurtzite content

GaAs nanowires with ~30% wurtzite structure were synthesized using lower growth rates [51]. Representative HRTEM measurements of these nanowires are presented in Fig.6, where a bright field TEM image is shown. The average length of the nanowires is about 3 μm. From the structural point of view, it presents two distinctly different regions –labeled A and B in Fig.6a-. Region A is formed by nearly 100% wurtzite, as indicated by the analysis of the corresponding power spectrum of Fig.6f. Region B is dominated by the zinc-blende phase, and only in a narrow region close to region A, a mixture of the zinc-blende and wurtzite is observed. Further away from region A, the density of the zinc-blende twins decreases.

Polarization dependent Raman spectra have been measured along the nanowire with a spacing of 100 nm. The polarization dependence of the Raman scattering was characterized by measuring both the azimuthal dependence of the Raman scattering and the spatial dependence along the nanowire under the four main configurations -consisting of the incident and analyzed polarization being parallel or perpendicular to the nanowire axis z-. Such measurements were taken along the whole length of the nanowire. An overview of the measurements realized under the four main polarization configurations is presented in Fig.7 as color coded mapping scan. The scans can be divided into three different regions, presenting different features. In the lower part of the scan the $E_1^H$ (TO) mode is observed at 266.7 cm$^{-1}$, as expected. Furthermore, the $A_1$(LO) mode at 290.9 cm$^{-1}$ is present with weaker intensity. It is worth to note again that this mode is not allowed for the backscattering configuration on {110} family surfaces. The upper part of the nanowire is characterized by the presence of a further peak at about 256 cm$^{-1}$, which corresponds to the $E_2^H$ mode from the wurtzite GaAs phase. The $E_1^H$ (TO) mode is also observed. The $E_2^H$ mode is found at the upper 2.2 μm of the nanowire, the intensity decreasing towards the middle. The intensity evolution corresponds well with the percentage of wurtzite phase in the nanowire. As shown by the HRTEM analysis along the nanowire, the top part is composed mainly of wurtzite (98%). Then, towards the middle of the nanowire, there starts to be a mixture between wurtzite and zinc-blende. Zinc blende becomes dominant at the end of the nanowire. The middle part of the scan, which extends about 1μm, presents mainly just the $E_1$(TO) mode, showing typical features of zinc-blende structure. In the following, we first describe the characteristics of the spectra in the middle part of the nanowire. For clarity, we have marked the position in Fig.7a. In Fig. 8a we present the Raman spectra obtained under the four main polarization configurations. For comparison, in Fig.8b we present the equivalent measurements realized on a nanowire consisting of 100% zinc-blende structure, already described in the previous section. The corresponding peak positions and FWHM are listed in Table 2. The spectra taken under $x(y,z)\overline{x}$ configuration have been normalized and the others are plotted with the relative intensities with respect to the first one. Both samples show a strong $E_1$(TO) mode under the $x(z,z)\overline{x}$ configuration. The $A_1$(LO) mode is only weakly present in the spectra collected from the nanowire with 30% wurtzite. The variation in the scattered intensity as a function of the polarization configuration is significantly different between the two samples. As already discussed above, the intensity of the $E_1$(TO) mode for zinc-blende nanowire exhibits a drastic drop in the case where the exciting light is perpendicularly polarized. Interestingly, all the spectra collected from the nanowire presenting 30% of wurtzite structure exhibit a relatively intense $E_1$(TO) mode, with a much weaker dependence on polarization. This phenomenon may be explained by the presence of the high density of twins and a mixture between the wurtzite and zinc-blende structures. Indeed, the $E_1$(TO) mode in the wurtzite structure is allowed for crossed polarization from the selection rules, as given in Table 1. The FWHM of the $E_1$(TO) mode measured under the configuration $x(z,z)\overline{x}$ for the sample presenting 30% of wurtzite structure is 3.7 cm$^{-1}$, and thus one cm$^{-1}$ smaller compared to the spectra obtained from the nanowire consisting of 100% zinc-blende structure. The reason for this is still not yet clear. It may be related to the different size of the nanowire.

As shown in the Raman scans presented in Fig. 7, the $E_2^H$ mode, which is clearly visible only in the upper part of the wire, appears with the highest intensity in $x(y,y)\overline{x}$ configuration where the incident and analyzed polarizations are perpendicular to z. This is in good agreement with the selection rules summarized in Table 1. On the contrary, the $E_1(TO)$ mode shows much less selectivity, being observed for all polarizations. The intensity of $E_1(TO)$ is highest when the incident and analyzed polarization are parallel, especially when they are parallel to the z axis. This is consistent with what we have observed for the $E_1(TO)$ mode in pure zinc-blende nanowire.

In order to obtain a clearer understanding on the selection rules in nanowires containing wurtzite inclusions, the azimuthal dependence of the Raman scattering was measured in a region of a nanowire where wurtzite was majority. The azimuthal dependence is shown both for detected polarizations parallel and perpendicular to the z axis in Fig.9. As in the case of zinc-blende nanowires, the $E_1(TO)$ mode is polarized along the axis of the nanowire. Interestingly, also $I_s(\parallel)$ seems to have a higher intensity when the incident light is polarized along the nanowire axis. The ratio of intensity between $I_s(\parallel)$ and $I_s(\perp)$ is about 5. The azimuthal dependence of the $E_2^H$ mode associated with the wurtzite phase exhibits a quite different behavior. Indeed, in this case the maximum intensity of the scattered light is observed when the incident light is perpendicular to the nanowire axis, both for $I_s(\parallel)$ and $I_s(\perp)$ –though for the latter the dependence is less clear due to the low intensity-. Also the "forbidden" $A_1(LO)$ mode is observed in some cases. Due to the low intensity, the polarization dependence of this mode could not be determined with high enough accuracy. The possible origin of this mode will be discussed in section 5.3.

## 5. Discussion

### 5.1 Raman selection rules for nanowires

The azimuthal dependence of the Raman scattering in sections of the nanowire with zinc-blende structure demonstrate a tendency for the modes to be more efficiently excited for parallel polarization. We find that the intensity of the scattered modes is much higher in the case of emission with polarization parallel to the nanowire axis. As it was shown in Section 2, these two characteristics are different from what is expected in zinc-blende GaAs. Polarization dependent Raman studies on single GaP nanowire have demonstrated that nanowires present a dipolar or multipolar behavior as a function of the diameter [58]. The origin of this effect has been attributed to the scattering of the electromagnetic field from a dielectric cylinder of nanoscale dimensions. In the case of nanowires with a diameter $d >> \lambda_{laser} / 4$, the electric field inside the nanowire is enhanced when the electric field of the excitation is either parallel or perpendicular to the nanowire axis. Instead, for diameters $d << \lambda_{laser} / 4$, the electric field inside the nanowire is strongly suppressed when the electric field of the excitation is perpendicular to the nanowire axis. Furthermore, experiments on silicon nanocones showed that the enhancement in the Raman scattering, due to the enhanced internal field, decreases with the nanowires diameter and increases with the wavelength of the excitation, features which suggest a resonant nature [32]. This is consistent with luminescence experiments (PL). Indeed, it is observed that the light is preferentially absorbed when the incident light is polarized along the nanowire axis [59]. One distinct feature of nanowires embedded in an environment with a different dielectric constant is, that both absorption and emission for different polarizations of the light depend strongly on its frequency and the nanowire radius. This is due to image forces building up at the interface between the nanowire and its surrounding. Indeed, it has been shown that the field distribution is non-uniform inside the nanowire [60]. Absorption is not governed by the details of this field distribution, but by the total power absorbed, which is proportional to the light intensity integrated over the nanowire cross section. The result is that the polarization ratio will depend on the diameter of the nanowire. For the diameters measured in this paper, the polarization dependence of the absorption should oscillate between 60 and 90% [55,61]. In the case of wurtzite arsenide or phosphide nanowires, luminescence tends to be polarized perpendicularly to the nanowire axis. This is due to the symmetry of band structure, in which the recombination of electron-hole pairs is only allowed for an electric field perpendicular to the c axis of the nanowire [37,62,63]. Still, the dielectric contrast with the environment modulates the polarization dependence rendering it smaller.

Finally, we comment on the possible splitting of the TO and LO modes, giving rise to $E_1$(LO) and $A_1$(TO) modes. In cases where the splitting is very small, it is manifested by an asymmetrical broadening of the peaks in measurements where the polarization is not defined [64]. For polarization dependent Raman scattering, the two split modes should be present independently for parallel and perpendicular polarization. We have seen that the position of the $E_1$(TO) mode depends on the analyzed polarization. This can be seen for example in Fig. 9c and Table 2, where spectra and peak positions with different polarization configurations are shown. For the configurations $x(y,z)\bar{x}$ and $x(z,z)\bar{x}$, we find consistently in all measurements the $E_1$(TO) peak respectively at 267 and 265 cm$^{-1}$. This difference in the peak position occurs only in sections where wurtzite is present –for 100% zinc-blende nanowires it is not observed-. In the case of the $E_2^H$ mode, the splitting seems to be more important: about 4 cm$^{-1}$. No clear polarization dependence is found for the position of the $A_1$ mode. Two effects could explain the splitting: (i) the splitting of the $E_1$(TO) and $A_1$(LO) modes leading to the modes $E_1$(LO) and $A_1$(TO) and/or (ii) that the position of the $E_1$(TO) mode is not the same for the wurtzite and zinc-blende structure. A more detailed study on the splitting as a function of the nanowire crystal structure, strain, diameter and length is currently under way.

In conclusion, the selection rules of wurtzite and zinc-blende GaAs nanowires are dominated by the nanowire geometry and the symmetry of the electron-hole absorption. Moreover, uniquely in the nanowire regions where wurtzite is present, the $E_1$(TO) mode presents a polarization dependent splitting of ~2 cm$^{-1}$.

### 5.2 Effect of strain in the peak position

So far we have not considered the possible existence of strain in nanowires of wurtzite/zinc-blende structures. It is known that wurtzite exhibits a slightly larger lattice parameter than zinc-blende [65]. Raman spectroscopy is a simple technique to obtain information about local strain in materials, as built-in strain leads to a characteristic shift of the phonon frequency [66,67]. By realizing spatially resolved Raman spectroscopy, variations of the strain along the nanowire can be measured. Such variations should occur if the relative percentage of the two phases changes. Indeed, in regions where wurtzite dominates, the zinc-blende sections should be under tensile strain. Along the same line, wurtzite should be compressively strained in regions with mainly zinc-blende phase. In Raman spectroscopy, the tensile/compressive strains are characterized by a shift towards lower/higher wavenumbers.

A nanowire presenting a clear gradient in the percentage of wurtzite and zinc blende was investigated. The gradient was achieved by gradually changing the arsenic pressure during growth [51]. A representative bright field TEM micrograph of the first 1.5 μm of such nanowire is presented in Fig.10. The percentage of wurtzite varies from 98% at one extreme of the nanowire to 32% in the center and close to 0% after the 1.5 μm shown. This represents a strong gradient in a relatively short distance.

Spatially resolved Raman spectroscopy was realized on this type of nanowires. The measurement realized in $x(y,y)\bar{x}$ configuration is presented in Fig. 11 in a waterfall plot. Three peaks with changing intensity ratios are observed for all positions: $E_2^H$, $E_1^H(TO)$ and $A_1^H(LO)$. As can be clearly seen, the position of $E_2^H$ and $E_1^H(TO)$ shift along the scan. In order to provide a more detailed analysis, we plot the peak position, intensity ratio $E_2^H/E_1^H(TO)$ and the FWHM of the peaks $E_2^H$ and $E_1^H(TO)$ along the nanowire in Fig. 11c, d and e. The data presented are given by the average of the parameters extracted by the fits of spectra accumulated in three consecutive positions. The highest ratio between $E_2^H$ and $E_1^H(TO)$ is given at position 0.3 μm. This is in agreement with the TEM micrograph in which a 100% wurtzite section is found between 100 and 600 nm. At that point the position of the peak $E_2^H$ is 254 cm$^{-1}$, close to the value expected for wurtzite GaAs. At that position, the FWHM of $E_2^H$ is also the smallest, namely 4.5 cm$^{-1}$. The $E_1(TO)$ mode is found at 265.6 cm$^{-1}$ with a FWHM of 5.6 cm$^{-1}$. The downshift of the $E_1(TO)$ corresponds to a tensile strain of the zinc-blende structure, as expected. For positions away from 0.3 μm, the ratio $E_2^H/E_1^H(TO)$ decreases in agreement with a gradual increase in the zinc-blende phase –as it was observed by TEM-. Along with this gradual increase in zinc-blende, a shift of both the $E_2^H$ and $E_1^H(TO)$ modes towards higher wavenumbers appears. This shift to higher wavenumbers corresponds to a relaxation of the zinc-blende structure, with a consequent shift of the corresponding peak to its original value, and a compressive strain of the wurtzite structure. At the position 0.7 μm the energy of the $E_1^H(TO)$ mode has shifted to the expected value of 266.2 cm$^{-1}$. The percentage of zinc-blende at this position is about 68%. The percentage of zinc-blende further increases along the nanowire towards 100%. The wurtzite peak $E_2$ is still shifting and it stabilizes at ~1 μm. Consistently, we observe that the FWHM is larger at the positions where the respective crystalline phase is under strain.

The maximum shift observed both from the $E_1(TO)$ and the $E_2$ mode is about 4 cm$^{-1}$. One should note that the two modes are always shifted by the same amount, which is expected for small values of strain. The existence of strain in wurtzite/zinc-blende structures is in agreement with the slight difference in lattice constants of the two structures [68]. This means that for zinc-blende rich nanowires the wurtzite phases exhibit compressive strain (blue shift of the Raman mode) while for wurtzite rich nanowires the zinc blende phases are under tensile strain (red shift of the Raman modes). The existence of strain in singular heterostructures is not expected to exist in nanowires [69]. In this case the structure is not a simple heterostructure, but a mixture between two crystalline phases in which one of them is minority. In this specific case, the existence of strain is to be expected.

### 5.3 Effect of surface roughness on the appearance of $A_1(LO)$ modes

Finally, we would like to comment on the measurement of the $A_1^H(LO)$ mode at one end of the nanowire. For this, we have systematically looked at the faceting of the nanowires in the whole length. We present the example of the nanowires composed of 30% of wurtzite, where the intensity of the $A_1^H(LO)$ mode is always higher at the tip of the nanowire where the percentage of zinc-blende phase is the highest. A representative HRTEM micrograph of this part of the nanowire is presented in Fig. 13. The crystal structure is zinc-blende with twins separated between 3 and 10 nm. In the case where the twins are closer, one observes that the facets of the nanowire are not of the family {110} anymore, but {111}. The $A_1^H(LO)$ mode is allowed for backscattering from {111}. We believe that the small {111} faceting at the end of the nanowire leads to an increased intensity of the $A_1^H(LO)$ phonon mode.

## 6. Conclusion

In conclusion, we have presented spatially resolved Raman scattering experiments on single GaAs nanowires, with different percentages of zinc-blende and wurtzite structure. The polarization dependence and selection rules have been obtained for both crystalline phases. The TO mode related to the wurtzite structure, $E_2^H$, is found at 254 cm$^{-1}$, in agreement with the theoretical considerations. The azimuthal dependence of $E_2^H$ indicates that the mode is excited with the highest efficiency when the incident and analyzed polarization are perpendicular to the nanowire axis. In the case of zinc-blende, the transversal optical mode $E_1(TO)$ at 266.3 cm$^{-1}$ exhibits the highest intensity when the incident and analyzed polarization are parallel to the nanowire axis. This is a combined consequence of the selection rules and the one-dimensional geometry of the nanowire and the dielectric mismatch with the environment. The presence of strain in regions where either wurtzite or zinc-blende is in minority is measured by the shift of the $E_1(TO)$ and $E_2^H$ modes. Finally, we find that the existence of surface roughness can enable scattering by longitudinal optical modes on {110} facets.

*Acknowledgements*

The authors would like to kindly thank T. Garma, J. Dufouleur, B. Ketterer and M. Bichler for their experimental support and discussions. This research was supported by Marie Curie Excellence Grant 'SENFED', and the DFG excellence cluster Nanosystems Initiative Munich as well as SFB 631. The Swiss National Science Foundation is acknowledged for the funding of project "Catalyst-free direct doping of MBE grown III-V nanowires". The Authors would like to thank the TEM facilities in the Interdisciplinary Center for Electron Microscopy in Lausanne.

**Table 1.** Phonon modes in wurtzite GaAs. The activity in Raman and/or infrared absorption measurements is indicated, as well as the configurations for which Raman measurements are possible.

| Mode | Position (cm⁻¹) | Active | Configuration |
|------|------|------|------|
| $A_1$(LO) | 291 | Raman, Infrared | $z(y,y)\bar{z}$ |
| $A_1$(TO) | not known | Raman, Infrared | $x(y,y)\bar{x}$ , $x(z,z)\bar{x}$ |
| $B_1^H$ | 234 | Silent mode | - |
| $E_2^H$ | 259 | Raman | $x(y,y)\bar{x}$ , $x(y,y)\bar{z}$ , $z(y,x)\bar{z}$ , $z(y,y)\bar{z}$ |
| $E_1^H$ (TO) | 267 | Raman, Infrared | $x(z,y)\bar{x}$ , $x(y,z)y$ |
| $E_1^H$ (LO) | not known | Raman, Infrared | $x(y,z)y$ |
| $B_1^L$ | 206 | Silent mode | - |
| $E_2^L$ | 59 | Raman | $x(y,y)\bar{x}$ , $x(y,y)\bar{z}$ , $z(y,x)\bar{z}$ , $z(y,y)\bar{z}$ |

**Table 2.** Position and linewidth of $E_1$ and $E_2$ peaks for the bulk reference sample, nanowires (Nw) with 100% zinc-blende structure (ZB), 30% wurtzite in the region with a majority of zinc-blende and in the region with a majority of wurtzite (WZ).

| Mode | Sample | Position (cm⁻¹) | FWHM (cm⁻¹) | Configuration |
|------|--------|-----------------|-------------|---------------|
| $E_1$(TO) | Bulk GaAs (0-11) | 267.2 | 2.8 | $x(z,z)\bar{x}$ |
| $E_1$(TO) | Nw 100% ZB | 266.7 | 4.8 | $x(z,z)\bar{x}$ |
| $E_1$(TO) | Nw 30% WZ (ZB) | 267.1 | 3.7 | $x(z,z)\bar{x}$ |
| $E_1$(TO) | Nw 30% WZ (ZB) | 267.2 | 4.1 | $x(y,z)\bar{x}$ |
| $E_1$(TO) | Nw 30% WZ (ZB) | 266.9 | 3.7 | $x(y,z)\bar{x}$ |
| $E_1$(TO) | Nw 30% WZ (ZB) | 267 | 3.8 | $x(y,y)\bar{x}$ |
| $E_1$(TO) | Nw 30% WZ (WZ) | 264.9 | 5.9 | $x(z,z)\bar{x}$ |
| $E_1$(TO) | Nw 30% WZ (WZ) | 267 | 4.6 | $x(y,z)\bar{x}$ |
| $E_1$(TO) | Nw 30% WZ (WZ) | 265.8 | 5.5 | $x(z,y)\bar{x}$ |
| $E_1$(TO) | Nw 30% WZ (WZ) | 266.2 | 5.2 | $x(y,y)\bar{x}$ |
| $E_2^H$(TO) | Nw 30% WZ (WZ) | 256.1 | 4 | $x(y,y)\bar{x}$ |
| $E_2^H$(TO) | Nw 30% WZ (WZ) | 255.9 | 4.9 | $x(z,y)\bar{x}$ |

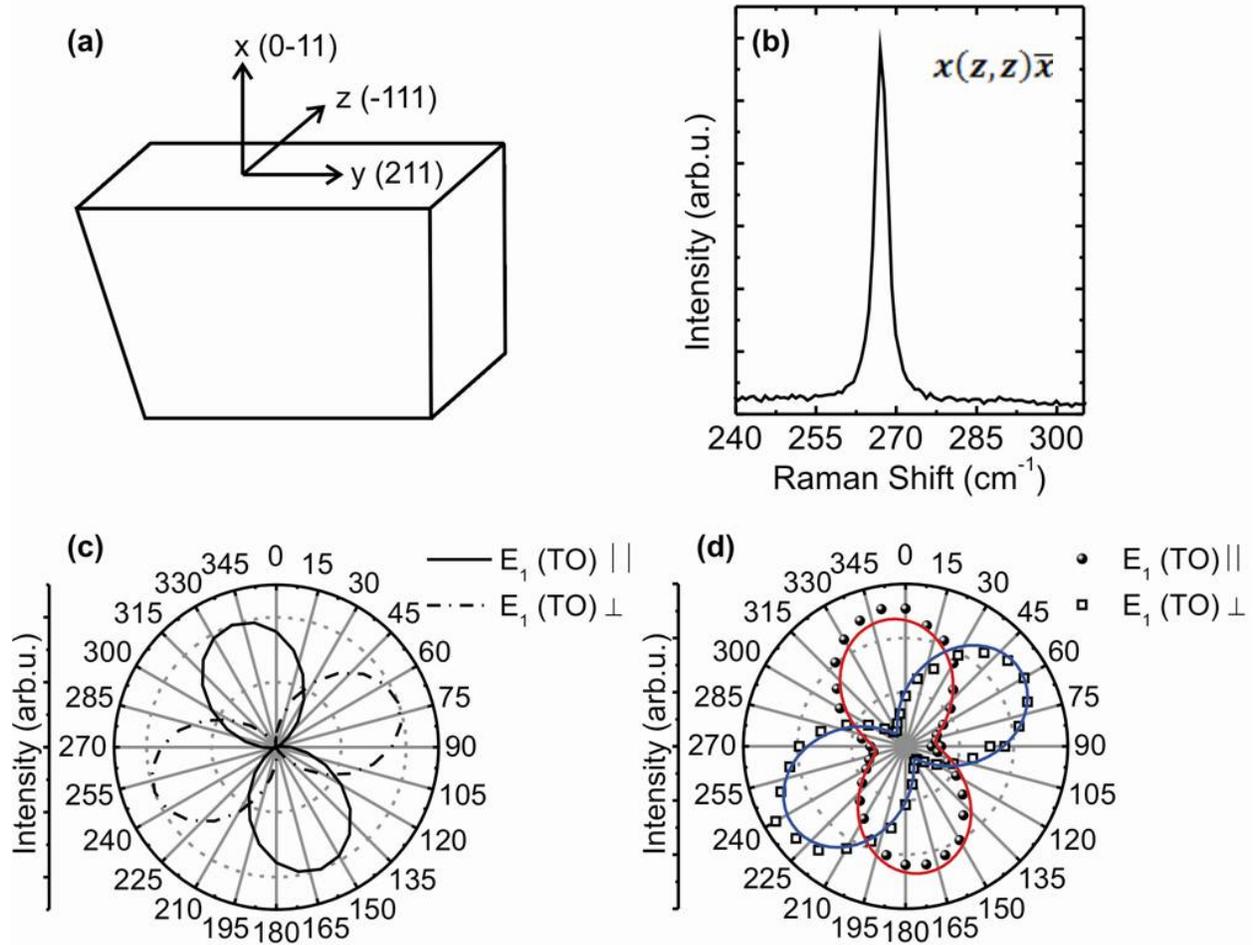

Figure 1. (a) Crystal facets of the reference used for the measurement of the selection rules in GaAs. The axis correspond to the crystallographic directions: x=[0-11], y=[211] and z=[-111]. (b) Parallel polarized Raman spectrum from a bulk GaAs (0-11) in backscattering geometry, obtained using exciting light polarized along the [-111] direction. (c) Theoretical azimuthal dependence of the TO mode of a bulk GaAs (0-11), as in (a). Continuous and dashed lines represent the components along the [-111] (ǁ) and [211] (⊥) of the Raman signal, respectively. (d) Measured azimuthal dependence of the TO mode of a bulk GaAs (0-11). Spheres and open squares represent the components along the [-111] (ǁ) and [211] (⊥) of the Raman signal, respectively. The continuous line is a fit to the data with a squared sine function, which describes dipolar behaviour.

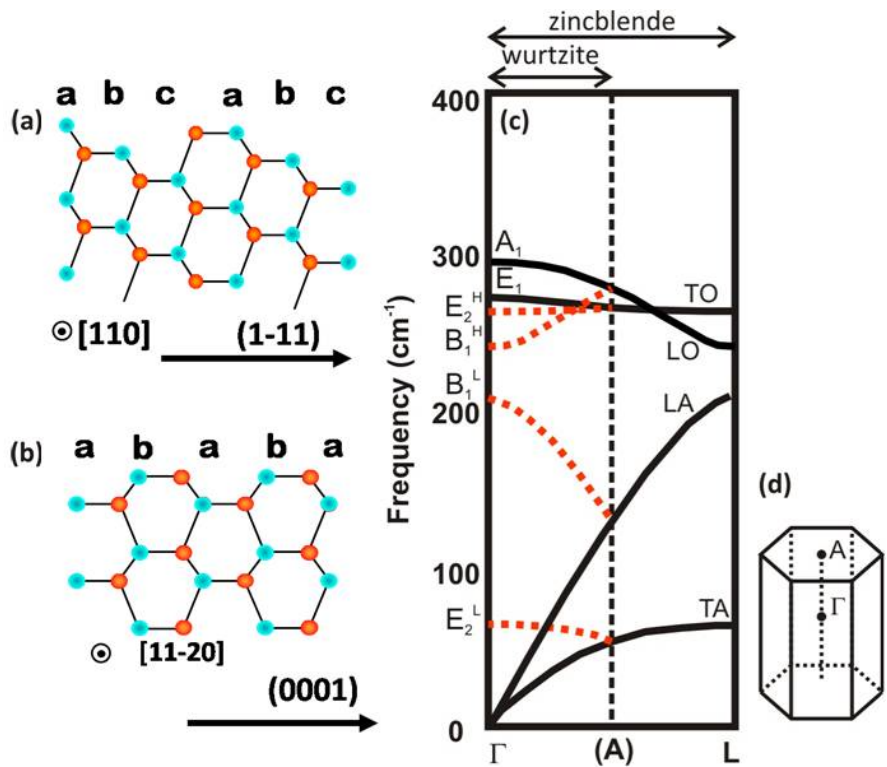

Figure 2. Schematic drawing of the atomic arrangement in zinc-blende **(a)** and wurtzite **(b)** structures, , where the arrows indicate the [1-11] and the [0001] nanowire growth axes, respectively. (c) Schematic representation of the phonon dispersion in GaAs. Phonon branches along [111] in the zinc-blende structure are folded to approximate those of wurtzite structure along [0001] (d) Unit cell for the wurtzite structure. Point A corresponds to the edge of the first Brillouin zone in direction [001].

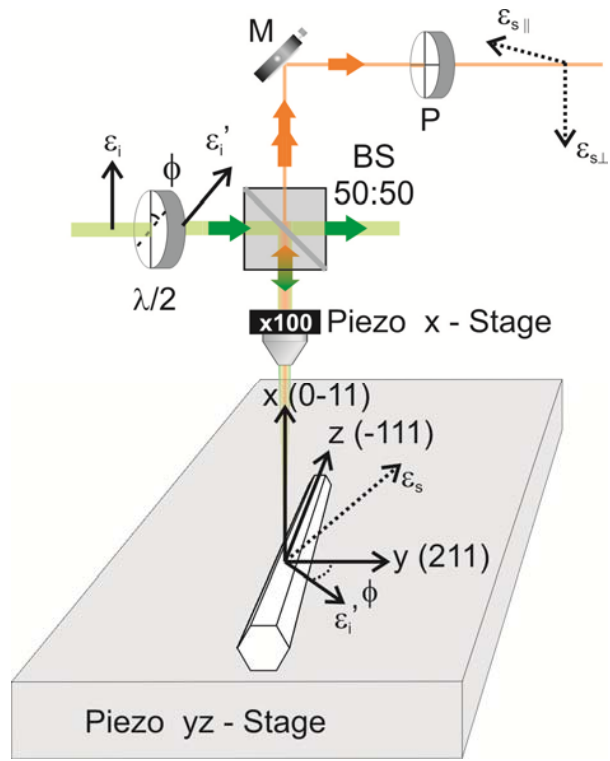

Figure 3. Sketch of the experimental setup and the configurations used for the measurement of GaAs nanowires in backscattering geometry. The crystal facets of the nanowire and the corresponding set of axis used as indicated: x=[0-11], y=[211] and z=[-111].

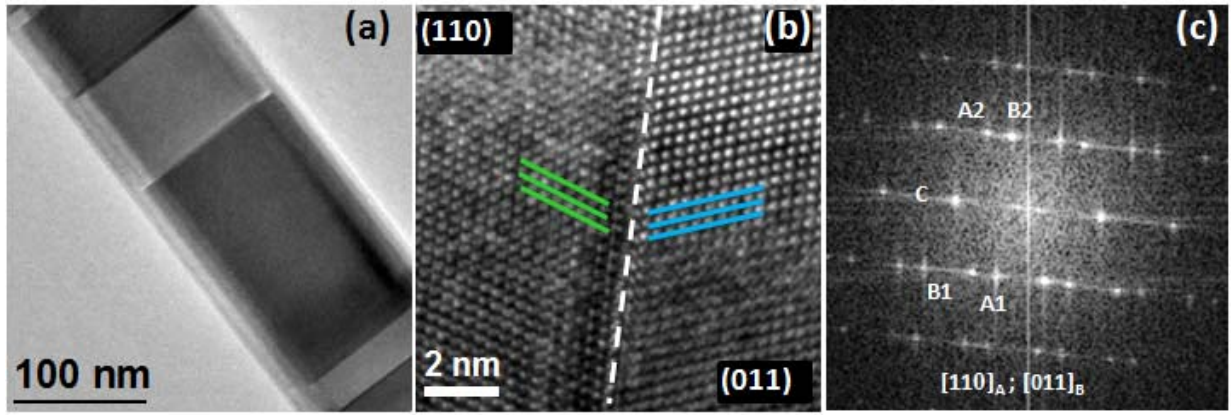

Figure 4. (a) BF image of a GaAs nanowire grown under conditions where 100% zinc-blende is obtained; (b)HRTEM of a twin in the zinc-blende structure (c) power spectrum of the twined region. The different spots can be identified with the two regions of the twin. The facets on axis are identified by (110) and (011).

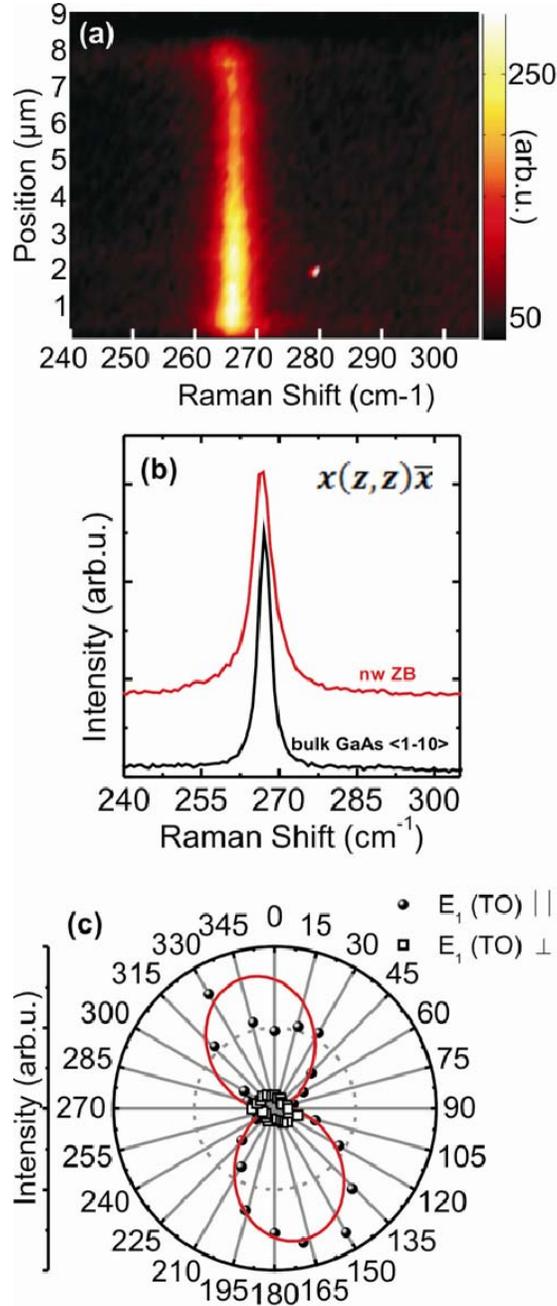

Figure 5. (a) Raman scan from a nanowire crystallized in ZB structure, obtained using exciting light polarized along the nanowire axis. (b) Parallel polarized Raman spectra from a bulk GaAs (0-11) and a GaAs nanowire crystallized in the ZB structure. For both measurements the exciting and scattered light are polarized along the <-111> direction. The spectrum of the nanowire is shifted vertically in order to make the comparison easier. (c) Azimuthal dependence of the TO mode related to the ZB structure in the nanowire. Spheres and open squares represent the parallel and perpendicular components of the Raman signal collected with respect to the nanowire axis, respectively. The continuous line is a squared sine fit to the data.

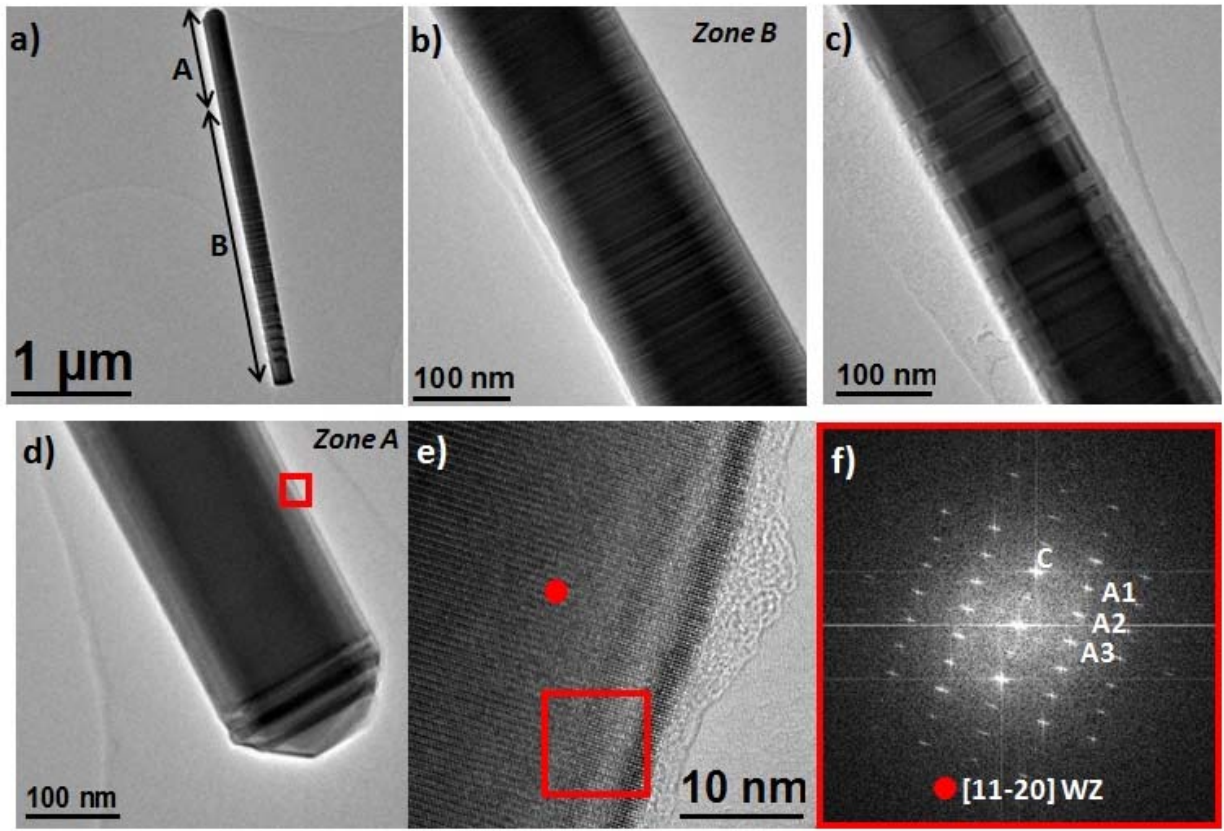

Figure 6. (a) BF image of a GaAs nanowire grown under conditions where ~30% wurtzite structure is obtained; (b)-(c) Region of multitwinned zinc –blende in the region B, (d) Bright field TEM image of the tip of region A (e) HRTEM in the selected square of (d), (f) power spectrum of the region marked with a square in e), which allows the identification of the crystalline structure of this region to be wurtzite, where spots C, A1, A2 and A3 correspond to (0001), (1-100), (1-101) and (1-102) planes, respectively.

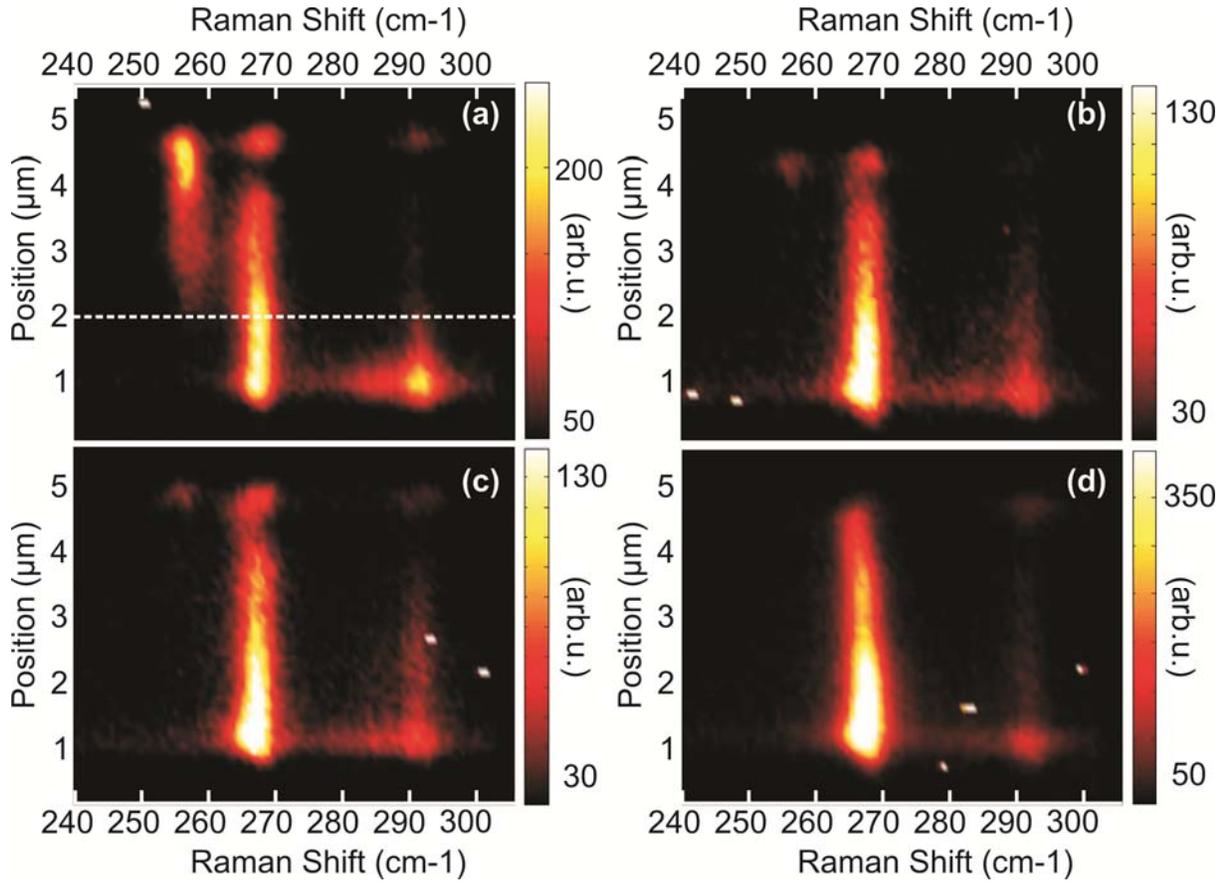

Figure 7. Perpendicularly and parallel polarized Raman scans from a nanowire consisting of 30% wurtzite structure, obtained using different polarization directions of the incident light: a) Perpendicularly polarized Raman scan from perpendicularly polarized incident light: $x(y,y)\bar{x}$ b) parallel polarized Raman from perpendicularly polarized incident light: $x(y,z)\bar{x}$ c) perpendicularly polarized Raman scan from parallel polarized incident light: $x(z,y)\bar{x}$ d) parallel polarized Raman scan from parallel polarized incident light: $x(z,z)\bar{x}$.

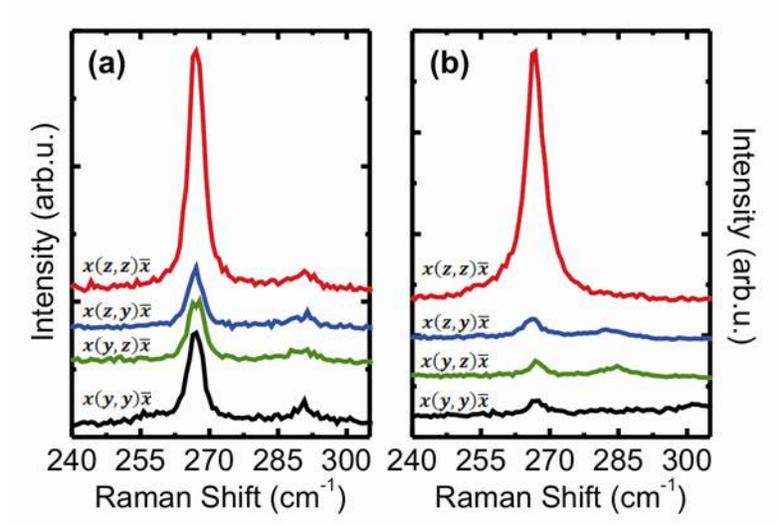

Figure 8. A series of parallel and perpendicularly polarized Raman spectra obtained using exciting light polarized parallel and perpendicularly to the nanowire axis, from (a) the zinc-blende rich segment of a nanowire with 30% of wurtzite structure and (b) a nanowire with 100% zinc-blende structure. The spectra have been shifted vertically. The relative intensities of the spectra obtained for the same sample in different scattering configurations can be compared directly.

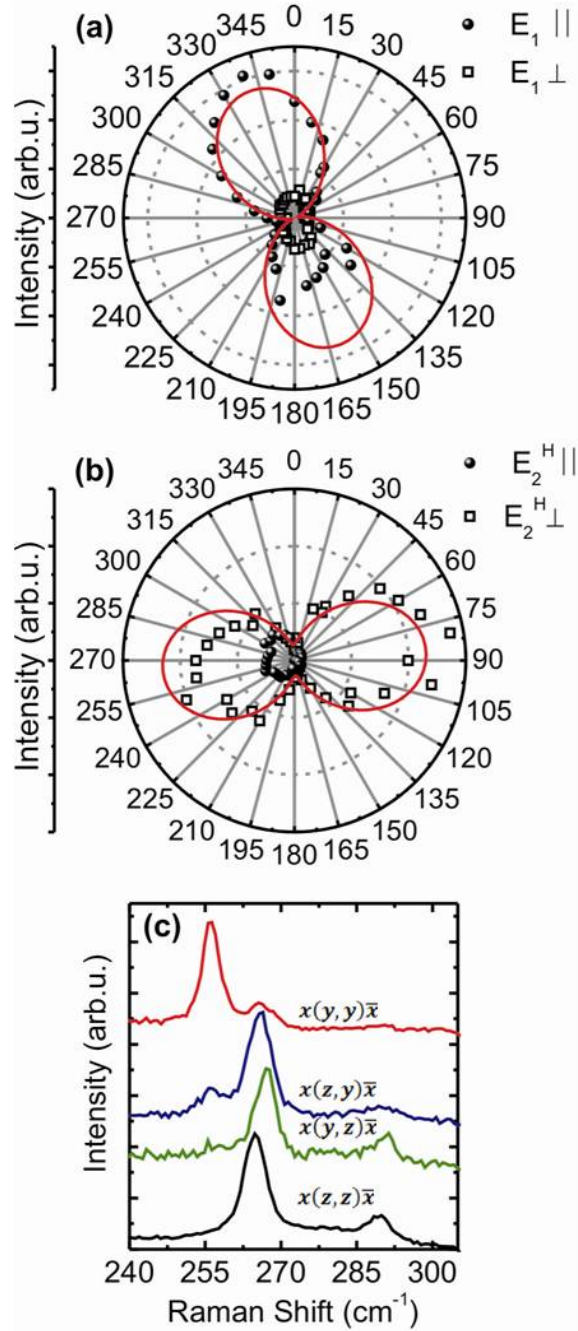

Figure 9. (a) Azimuthal dependence of the $E_1$(TO) mode. (b) Azimuthal dependence of the $E_2^H$ mode, related to the wurtzite structure. In both polar plots, spheres and open squares represent the parallel and perpendicular components of the Raman signal collected, respectively and the continuous line is a squared sine fit to the data. (c) Representative Raman spectra realized under the main four configurations. For better illustration, the spectra have been normalized and shifted vertically. All spectra have been realized in the same position of the nanowire.

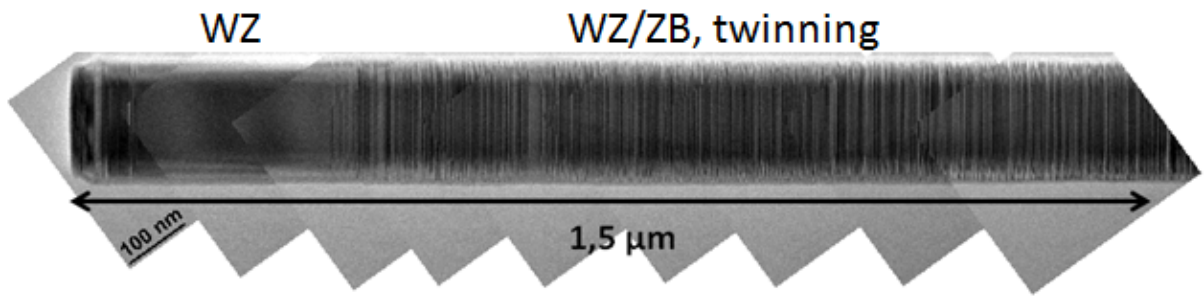

Figure 10. Bright field TEM measurement of the nanowire where one observes both the wurtzite and zinc-blende crystal structures, with a clear gradient of zinc-blende phase from left to right.

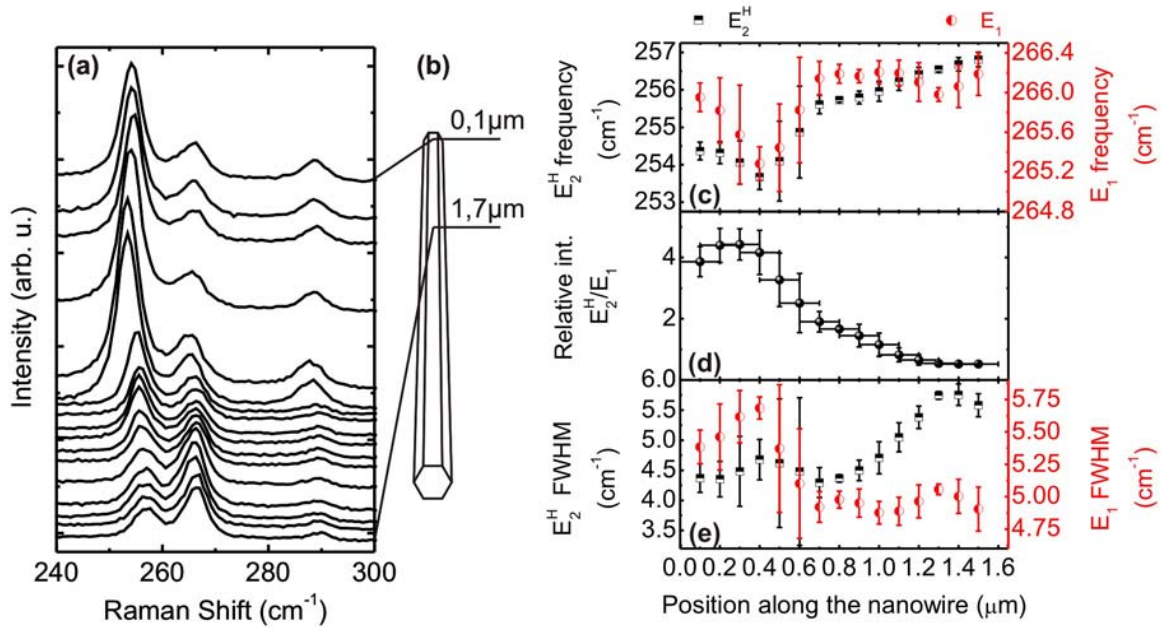

Figure 11. (a) Waterfall plot of Raman spectra collected along the nanowire, every 100 nm, covering a distance of 1.7 μm, as schematically depicted in (b); $E_2^H$ and $E_1$ peak positions and FWHM are plotted as function of the position along the nanowire in (c) and (e), respectively. (d) Relative intensity of the wurtzite and zinc-blende structure –calculated as the intensity ratio $E_2^H /E_1$ – as function of the position along the nanowire. Each of the points constitutes an average of three consecutive measurements.

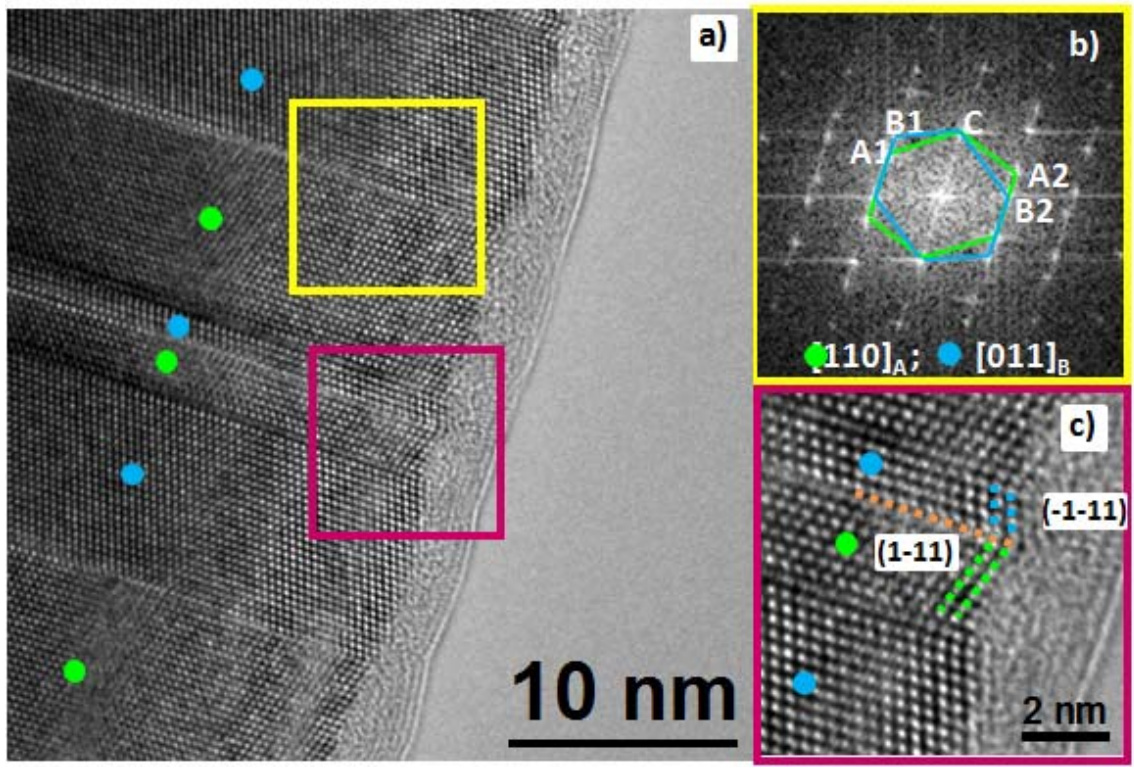

Figure 12. (a) HRTEM measurements close to the tip of a nanowire where the "forbidden LO phonon mode is observed; (b) power spectra of the region marked with yellow square in (a), where the diffraction spots of the two orientations of the twins are observed –sub-labels 1 and 2-, (*c*) Detail corresponding to the red square in the HRTEM micrograph, in which one can recognize the formation of (-1-1-1) facets.